# Transformational Plasmon Optics

*Yongmin Liu* [1], *Thomas Zentgraf* [1], *Guy Bartal* [1], *Xiang Zhang* [1,2,*]

[1]NSF Nanoscale Science and Engineering Center (NSEC), 3112 Etcheverry Hall,

University of California, Berkeley, CA 94720, USA

[2]Materials Science Division, Lawrence Berkeley National Laboratory,

1 Cyclotron Road, Berkeley, CA 94720, USA

*To whom correspondence should be addressed. E-mail: xiang@berkeley.edu

ABSTRACT

Transformation optics has recently attracted extensive interest, since it provides a novel design methodology for manipulating light at will. Although transformation optics in principle embraces all forms of electromagnetic phenomena on all length scales, so far, much less efforts have been devoted to near-field optical waves, such as surface plasmon polaritons (SPPs). Due to the tight confinement and strong field enhancement, SPPs are widely used for various purposes at the subwavelength scale. Taking advantage of transformation optics, here we demonstrate that the confinement as well as propagation of SPPs can be managed in a prescribed manner by careful control of the dielectric material properties adjacent to a metal. Since the metal properties are completely unaltered, it provides a straightforward way for practical realizations. We show that our approach can assist to tightly bound SPPs over a broad wavelength band at uneven and curved surfaces, where SPPs would normally suffer significant scattering losses. In addition, a plasmonic 180° waveguide bend and a plasmonic Luneburg lens with practical designs are proposed. It is expected that merging the unprecedented design flexibility based on transformation optics with the unique optical properties of surface modes will lead to a host of fascinating near-field optical phenomena and devices.

The recently developed technique of transformation optics (TO) provides a powerful means to precisely control the light flow in almost arbitrary ways [1-4]. Although the functionality of individual transformed optical devices varies from case to case, the methodology of TO is general. First, we distort part of free space in some manner that can be tracked by a certain coordinate transformation. Then the permittivity and permeability distribution in the new space can be obtained, taking advantage of the invariance of Maxwell's equations under coordinate transformations. As a result, light will propagate along desired trajectories. Normally the transformation results in a strongly inhomogeneous medium, that is, both the permittivity and permeability vary spatially and independently, with a large range of values throughout the transformed space. Metamaterials allow to tailor material properties beyond natural limitations by introducing subwavelength artificial structures [5,6], thus offering a wide range of accessible material properties to implement transformation optical designs. Indeed, numerous remarkable optical devices, such as invisible cloaks [7], beam rotators [8] and omnidirectional retroreflectors [9] have been demonstrated.

However, the material properties of the transformed space are in general highly anisotropic [7-11], requiring dispersive, resonant metamaterials. This significantly constrains the realization and application for practical devices, and results in disadvantages of narrow operation bandwidths and high losses. Some new strategies have been proposed to mitigate these limitations. For example, invisibility cloaks can be realized based on the scattering cancellation between a dielectric object and a metallic coating layer [12], or imitated by light propagation in specially tapered waveguides [13]. Also broadband waveguide bends utilizing gradient index metamaterials have been recently demonstrated [14]. In particular, the quasi-conformal mapping method has been developed to design carpet cloaks, broadband metamaterial lenses and optical Janus devices, in which only a modest range of isotropic refractive indices is required [15-20].

In principle, transformation optics embraces all forms of electromagnetic phenomena on all length scales. However, compared with the extensive work of transformation optics for propagation waves, much less considerable attention has been paid to near field optical waves, for example, surface plasmon polaritons (SPPs). SPPs arise from the collective electron density oscillation coupled to external electromagnetic waves [21]. The intensity of SPPs is maximal at the interface between a metal and a dielectric medium, while exponentially decay away from the interface. Due to the strong confinement and large field enhancement, SPPs are widely used for subwavelength imaging, lithography, novel optical devices and biomedical sensing [22,23]. The ability to control SPP propagation following certain transformation optical routes will undoubtedly lead to a variety of intriguing plasmonic devices with unprecedented performances and functionalities. Indeed, a few attempts were undertaken toward this goal. For instance, Smolyaninov *et al.* reported guiding and cloaking SPPs by alternatively stacking plasmonic modes with positive and negative group velocities [24]. Elser and Podolskiy proposed to eliminate parasitic scattering of SPPs using anisotropic metamaterials [25]. Very recently, based on a two-dimensional scalar model rather than the TO framework, Baumeier *et al.* theoretically investigated cloaking from SPPs utilizing the interference of the fields scattered by specially arranged individual scatters [26]. Nevertheless, these works demand precise controls over the structure geometries and/or material properties, considerably limiting the working bandwidth.

In this paper, applying the TO technique, we demonstrate that SPPs can be molded more effectively and efficiently than ever, with only isotropic, nonmagnetic and non-dispersive dielectric materials introduced into the transformed space. SPPs are bounded surface waves at metal-dielectric interfaces, which implies that both the metal and dielectric materials need to be transformed if rigorously following the TO approach. In practice, it is very difficult and challenging to spatially modify the metal property at the deep subwavelength scale. On the other hand, in the frequency region apart from the surface plasmon resonance frequency, which is of interest in most cases because of reasonably long propagation length of SPPs, a significant portion of SPP energy resides in the dielectric medium. For instance, more than 95% of the total energy of SPPs at an air-silver interface is carried in the dielectric for wavelengths larger than 400 nm. Since the most energy is carried in the evanescent field outside the metal, it is rational to control SPPs by only modifying the dielectric material based on the transformation optics technique, while keeping the metal property fixed. More importantly, we show that the transformed dielectric materials can be isotropic and non-magnetic, if a prudent transformation scheme is taken. Full-wave simulations of different transformed designs prove the proposed methodology. We demonstrate that SPPs can be tightly confined at curved surfaces without scarifying significant scattering losses over a broad wavelength region. Furthermore, we show that a 180° plasmonic bend with almost perfect transmission can be achieved by this technique. Finally, we demonstrate that transformation optics can provide a tool to modify the in-plane propagation of SPPs at the example of a plasmonic Luneburg lens. Our results manifest the novel applications of transformation optics in plasmonic systems.

The surface topological variation can modify the propagation characteristic of SPPs and lead to scattering of the energy into the far field. In addition to the intrinsic Ohmic losses of metals, scattering can be a major loss factor that limits the propagation length of SPPs [21]. Scattering exists when two plasmonic modes are mismatched, that is, they have different penetration depths into the dielectric and metal. Such mismatch happens for any discontinuity in geometries or material properties. For example, at a single boundary between two surface elements with different mode indices, typically 10%-30% of the SPP energy scatters into free space [25,27]. Recently, it was theoretically demonstrated that properly designed anisotropic metamaterials can completely suppress scattering of surface waves on a planar interface by matching the plasmonic mode profiles in the two separate regions [25]. Here, we show that scattering of SPPs on uneven and irregular surfaces can be dramatically minimized based on the TO approach. Therefore, the inherent two-dimensional optical wave feature of SPPs is preserved, and the propagation distance of SPPs is improved. More importantly, only isotropic, nonmagnetic dielectric materials are introduced into the structure design.

We start by considering SPPs propagating on an air-silver interface with a single protrusion as schematically shown in Fig. 1(a). The topology follows the function $x(z) = 0.2 \mu m \cdot \cos^2(z \cdot \pi / 2 \mu m)$ in the region $-1\mu m \le z \le 1\mu m$, whereas $z$ is the propagation direction and $x$ normal to the surface. Using a commercial finite-element analysis software (Comsol Multiphysics 3.5), we perform two-dimensional simulations to study the scattering of SPPs by the protrusion. The permittivity of silver is described by the Drude model $\varepsilon_m(\omega) = \varepsilon_\infty - \frac{\omega_p^2}{\omega(\omega + i\gamma_c)}$, where the high-frequency bulk permittivity $\varepsilon_\infty = 6$, the bulk plasmon frequency $\omega_p = 1.5 \times 10^{16} \, rad/s$, and the collision frequency $\gamma_c = 7.73 \times 10^{13} \, rad/s$ are obtained by fitting the experimental data from the literature [28]. Figure 1(b) depicts the magnetic field component of electromagnetic waves at 633 nm wavelength, in which SPPs are launched from the left boundary, and then pass the protrusion. From Fig. 1(b), one can see that the protrusion gives rise to a clear forward scattering into free space. About 26% of the SPP energy is radiated to the far field by this scattering process. This is a fairly big loss, considering that the energy attenuation due to the Ohmic loss is only about 4% for SPPs propagating the same lateral distance.

Transformation optics enables to create a virtual space, so that SPPs appear to propagate on a flat surface even though a protrusion physically exists. Consequently, the unfavorable scattering can be suppressed. In fact, this scenario is closely related to the carpet cloak demonstrated recently [15-18]. For a carpet cloak, a curved reflecting surface behaves like a flat reflecting surface because the beam profile and phase front of a reflected beam is undisturbed, no matter from which angle a beam is illuminated to the surface. Instead of the quasi-conformal method based on grid optimizations [15-18], a two-step procedure embedded in the Comsol Multiphysics environment is adopted to calculate the dielectric constant in the transformed space. In the following we highlight the key points of this convenient and efficient method, while details can be found in Refs. [29] and [30]. First, we numerically solve the inverse Laplace's equation, that is,

$$\left(\frac{\partial^2}{\partial x_1^{'2}} + \frac{\partial^2}{\partial x_2^{'2}} + \frac{\partial^2}{\partial x_3^{'2}}\right) x_i = 0, \ i = 1, 2, 3 \qquad (1)$$

in the transformed space, with proper combinations of Dirichlet and Neumann boundary conditions. **x** and **x'** are the coordinates in the original and transformed space, respectively. The solution of Equation (1) is a smooth function that represents the deformation field associated with the space transformation. The material parameters in the transformed space can be subsequently determined by

$$\varepsilon' = A\varepsilon A^{\mathrm{T}}/\det(A) \qquad \text{2(a)}$$

$$\varepsilon' = A\mu A^{\mathrm{T}}/\det(A) \qquad \text{2(b)}$$

where $\varepsilon$ and $\mu$ are the permittivity and permeability in the original space (with flat surface), respectively. $A$ is the Jacobian matrix with components defined as $A_{ij} = \partial x_i' / \partial x_j$. The Jacobian matrix characterizes the geometrical variation between the original space and the transformed space. It has been proved that the procedure described above is essential to minimize the Winslow functional. The calculated transformed material parameters are almost identical to those obtained from quasi-conformal mapping by minimizing the modified Liao's functional [30], except in very complex cases where Winslow functional may result in grid folding.

Following the above procedure, we have calculated the refractive index as well as the coordinate grids in the transformed space [Fig. 1(c)]. As Neumann boundary conditions are applied to the top and bottom boundaries, the grids can slide along the boundaries to realize nearly orthogonal, square grids. This feature is consistent with the quasi-conformal mapping. We obtain a refractive index profile that is isotropic and ranges from 0.81 to 1.39 without extreme values. After applying the refractive index profile on top of the metal surface with the bump, we expect that SPPs can smoothly pass the bump with negligible scattering loss as illustrated in Fig. 1(d). That is, the surface appears virtually flat for the SPPs, although physically the surface protrusion exists. This prediction is confirmed by the two-dimensional simulation [Fig. 1(e)]. For practical purposes we would prefer the refractive index of the transformed dielectric material larger than unity. This can be done by increasing the background refractive index. For instance, if we deal with SPPs originally on a $SiO_2$-Ag interface rather than on an air-Ag interface, the refractive index of the transformed space ranges from 1.17 to 2.02.

Compared with previous work [24-26], one major advantage of the spatial transformation of the plasmonic structure is the broadband performance, since solely isotropic and non-dispersive materials are used to realize the transformed dielectric material. We numerically calculate the scattering loss of the SPPs propagating for the geometry shown in Figs. 1(b) and 1(e) for wavelengths from 450 nm to 850 nm. Before the transformation, i.e., for the bare air-Ag interface, the scattering loss is between 14% and 43%. For shorter wavelengths, the protrusion is more pronounced with respect to the penetration depth of SPPs into the dielectric part. Consequently, the scattering loss gradually increases as wavelength decreases. In a striking contrast, the scattering loss of the SPPs is below 4.5% over the entire wavelength region once the transformed dielectric cladding is applied, confirming the broadband functionality of the design [Fig. 1(f)].

The shown concept of transforming the surrounding dielectric to generate a virtually flat surface can also be applied for irregular surface profiles. Hence, the scattering of SPPs can be efficiently suppressed even if irregular surface defects exist, which are demonstrated in Figs. 2(a)-(c). The scattering loss for the specific geometry in Figs. 2(a) is about 88% at the wavelength of 633 nm, severely hindering the propagation of SPPs. Remarkably, if a proper dielectric cladding is applied, the transmitted SPPs can be greatly increased to 95% as shown in Fig. 2(c). The slight scattering mainly arises from the complexity of the geometry and consequent rapid variation of the dielectric constant. We would like to emphasize that the method used here is very fast and efficient for calculating the transformed material properties when the geometry is arbitrary and complex, without suffering potential convergence and long optimization time in traditional grid generation processes.

We further amplify the surface topology to form a 180° bended dielectric-metal surface as shown in Fig. 3(a). In this case, almost all energy leaks to free space. The radiation is attributed to the fact that below a critical radius $r$, the lateral component of the SPP wave vector becomes smaller than the photon momentum in free space [31]. Therefore, the radial component of the $k$-vector becomes real and the electromagnetic fields can no longer be confined. Such a strong radiation loss significantly impedes the realization of plasmonic bends that are based on a single dielectric-metal interface, in contrast to widely investigated metal-dielectric-metal SPP bends [32]. However, we can transform the dielectric constant in the bend region to realize almost perfect transmission of SPPs at the curved dielectric-metal interface. There are many different ways to engineer the material properties of the bend, depending on which specific transformation is performed [33-35]. Here we use one simple design, whose refractive index of the dielectric part in the bend structure is given by [14, 33]

$$n = C/r \qquad (3)$$

with *C* denoting an arbitrary constant. Figure 3(b) shows the distribution of the refractive index when *C* =2.3 μm. This value is chosen in order to minimize the impedance mismatch of SPPs at the bend entrance and exist. Apparently, the material after transformation is isotropic with a gradient index change. Figure 3(c) shows the magnetic field of SPPs when the silver bend is covered by the designed dielectric material. SPPs propagate nearly without radiation loss around the bend and reach the bottom side of the metal slab. We calculate the transmission of the SPPs through the bend over a broad wavelength range, which exhibits almost perfect transmission [Fig. 3(d)]. In this simulation, we neglect the intrinsic loss of silver in order to focus on the radiation loss. In comparison, the transmission of SPPs for the bend without the transformed dielectric cladding is extremely low, especially in the longer wavelength region.

The transformed dielectric material in the bend region can be further simplified for practical purposes. Since the penetration depth of SPPs into the dielectric region is finite, the transformed region can be appropriately truncated to avoid extremely small refraction indices far away from the surface. Furthermore, multilayers with suitable dielectric constants can replace the continuous gradient-index materials. Figure 4(a) shows the refractive index distribution of the proposed SPP bend with the simplified design. The dielectric background of the straight waveguide region is silicon dioxide with refractive index 1.45. The curved metal strip is covered by five 200-nm-thick layers, whose refractive indices linearly change in steps of 0.145 from 1.595 to 1.015. Although the refractive index profile is approximated by only five different values, the performance only slightly suffers from the simplified structure [Figure 4(b)].

The transformation of the optical space provides an elegant means to modify the propagation characteristics of SPPs. However, the resulting index has to be spatially tailored, by structuring the dielectric material or mixing different types of nano-particles at the deep wavelength scale [17-19, 25]. This makes the practical realization challenging, especially at visible wavelengths. Instead of spatially modifying the refractive index of the dielectric material, the thickness of a homogeneous dielectric cladding layer can be varied to change the effective mode index of SPPs. It provides an alternative method to realize transformed plasmonic structures. Consider a dielectric/dielectric/metal structure with permittivities $\varepsilon_1/\varepsilon_2/\varepsilon_m$, the dispersion relation of SPPs is implicitly given by

$$\tanh(k_2 \varepsilon_2 d) = -\frac{k_1 k_2 + k_2 k_m}{k_2^2 + k_1 k_m}. \tag{4}$$

Here, $d$ is the thickness of the sandwiched dielectric layer, and $k_{1(2,m)} = \frac{\sqrt{\beta^2 - \varepsilon_{1(2,m)} \omega^2/c^2}}{\varepsilon_{1(2,m)}}$ with $\beta$ representing the SPP wave vector along the propagating direction and $c$ representing the speed of light in vacuum, respectively. When $d$ is very small, SPPs approach the behavior of SPPs for an $\varepsilon_1/\varepsilon_m$ single interface; while if $d$ is larger than the penetration depth of SPPs, the scenario is close to SPPs for an $\varepsilon_2/\varepsilon_m$ interface. As a result, we can readily change the effective mode index of SPPs, defined as $n_{eff} = \beta/k_0$, by tapering the thickness of the middle dielectric layer. Based on Eq. (4), we calculate the effective mode index of SPPs versus the dielectric layer thickness $d$ at 633 nm wavelength. In the simulation, we assume $\varepsilon_1=1$ and $\varepsilon_2= 2.56$. From Fig. 5(a), one can see that the mode index changes from 1.04 to 1.45 when $d$ gradually increases to 250 nm. Such an index range is sufficient for designing particular plasmonic devices.

As a proof-of-concept example, we numerically demonstrate a plasmonic Luneburg lens utilizing the tapered dielectric-dielectric-metal structure. The original Luneburg lens is a spherically symmetric gradient index lens which can focus plane waves to a perfect geometric point on the opposite side of the sphere [36]. Due to this unique property, Luneburg lenses have been used for commercial radar reflectors and satellite antennas. The refractive index of a Luneburg lens satisfies $n(r) = \sqrt{2-(r/R)^2}$, where $R$ is the radius of the lens. In the following, we use a dielectric cone structure to implement the plasmonic counterpart of a Luneburg lens, which is schematically shown in Fig. 5(b). As a specific design, we choose the radius of the cone base equal to 2 μm and the height of the cone equal to 200 nm. For a cone made of a dielectric material whose refractive index equals 1.4, the effective index of SPPs at 633 nm wavelength matches the index profile required by the Luneburg lens quite well. Three-dimensional full-wave simulations confirm the performance of the plasmonic Luneburg lens design. SPPs on the air-silver interface are focused on the perimeter of the cone base [Figs. 6(a) and 6(b)]. The slight radiation into free space can be further reduced by enlarging the radius of the cone, so that SPPs propagate through the dielectric cone more adiabatically. We also perform a two-dimensional simulation, taking into account the effective mode index of SPPs [Fig. 6(c)]. The result agrees well with the three-dimensional simulation, and substantiates our design methodology. We would like to point out that the tapered plasmonic structure could be fabricated by the gray-scale lithography technique. It is expected that other plasmonic optical elements [37], including beam shifters, beam splitters and concentrators, could be realized by the tapered plasmonic structure.

In summary, we have demonstrated the possibility to control SPPs in a prescribed manner with practically achievable materials. Since most of the SPP energy is carried in the evanescent wave outside the metal, we propose to mold the propagation of SPPs by modifying the dielectric material based on the transformation optics approach. Such a concept has been proved to be valid and effective through numerical simulations. We have shown the suppression of SPP scattering on an uneven surface, almost perfect transmission of SPPs in a bend structure, as well as focusing of SPPs by a planar Luneburg lens. These findings reveal the power of the transformation optics technique to manipulate near-field optical waves. We expect that many other intriguing plasmonic devices will be realized based on the methodology introduced here.

Note: As we completed this manuscript, we accidentally found that another group also completed a paper independently at the same time on a similar topic [38].

ACKNOWLEDGMENT

This work was supported by AFOSR MURI program (FA9550-04-1-0434) and NSF NSEC (DMI0327077).

FIGURES

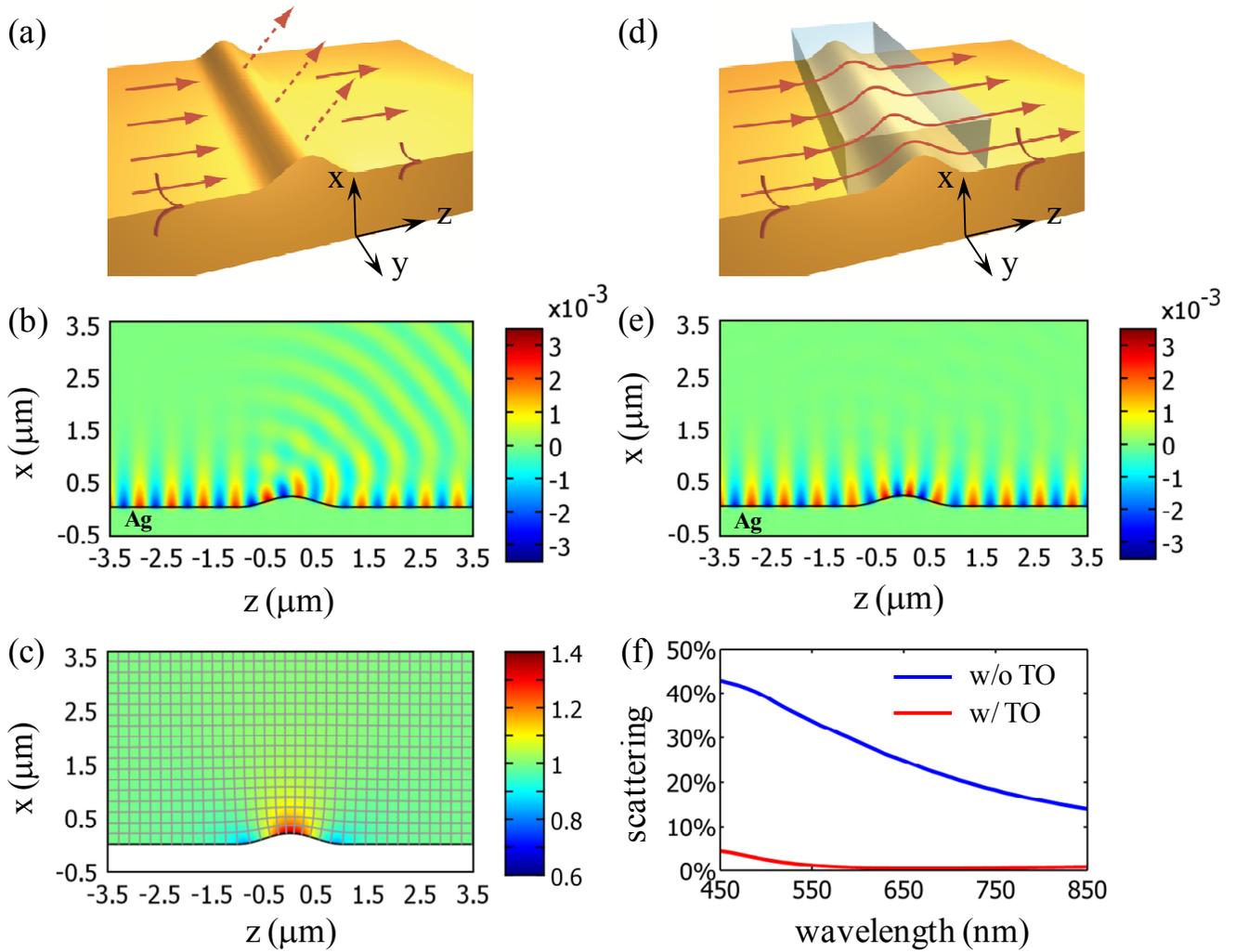

Figure 1. (a) Schematic of the SPP propagation and scattering on a metal-dielectric interface with a single protrusion. (b) Magnetic field component of electromagnetic waves when SPPs propagate on a protruded silver surface at 633 nm wavelength. Partial SPPs are scattered to free space. (c) The refractive index of the dielectric material in the transformed space, together with the transformed spatial grid. (d) Schematic of the SPP propagation when the dielectric space around the protrusion is transformed and the scattering is dramatically suppressed. (e) Magnetic field component of SPPs after the refractive index profile shown in (c) is applied to the dielectric space. (f) Scattering loss (the ratio of scattered energy and incident energy) of SPPs by the protruded surface at different wavelengths. The red line and blue line show the case with and without the transformation, respectively.

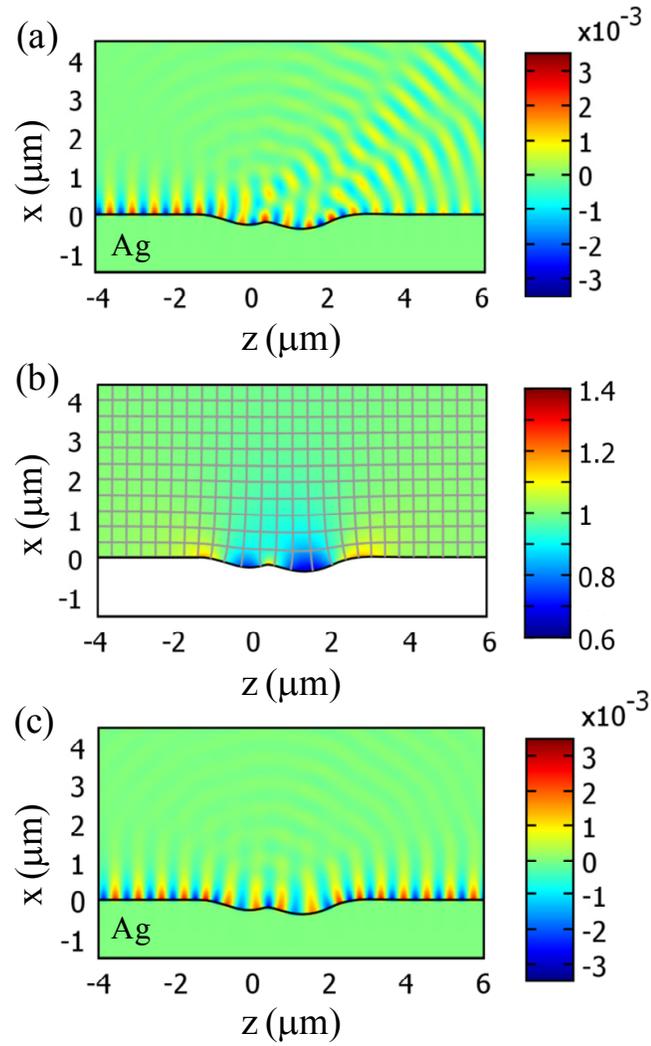

Figure 2. (a) Magnetic field component for SPPs propagating along surface with irregular surface modulations at 633 nm wavelength. (b) The refractive index distribution in the transformed space, together with the transformed spatial grid. (c) Same as (a) but after transformation of the dielectric space, showing significantly reduced scattering.

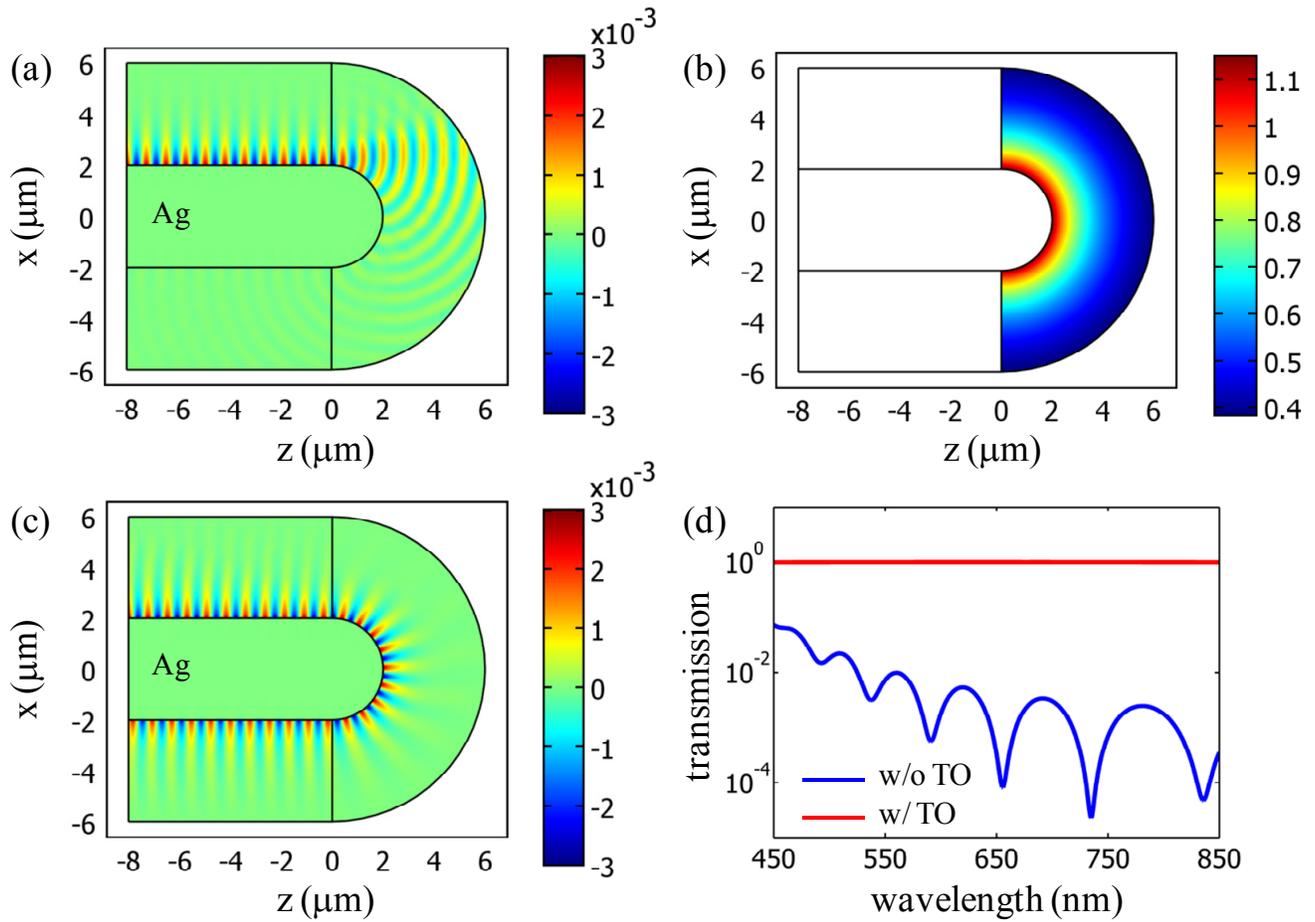

Figure 3. (a) SPPs propagation on a 180° bended dielectric-metal interface at 780 nm wavelength. Shown is the magnetic field component. (b) The refractive index of the transformed dielectric material in the bend region, which is inversely proportional to the radius of the bend. (c) Field distribution after the transformation of the dielectric material, exhibiting almost perfect transmission of SPPs in the bend structure. (d) Wavelength dependence of SPPs transmission of the bend structure (in a logarithmic scale). The red line and blue line represent the case with and without the transformation, respectively. The metal loss is neglected.

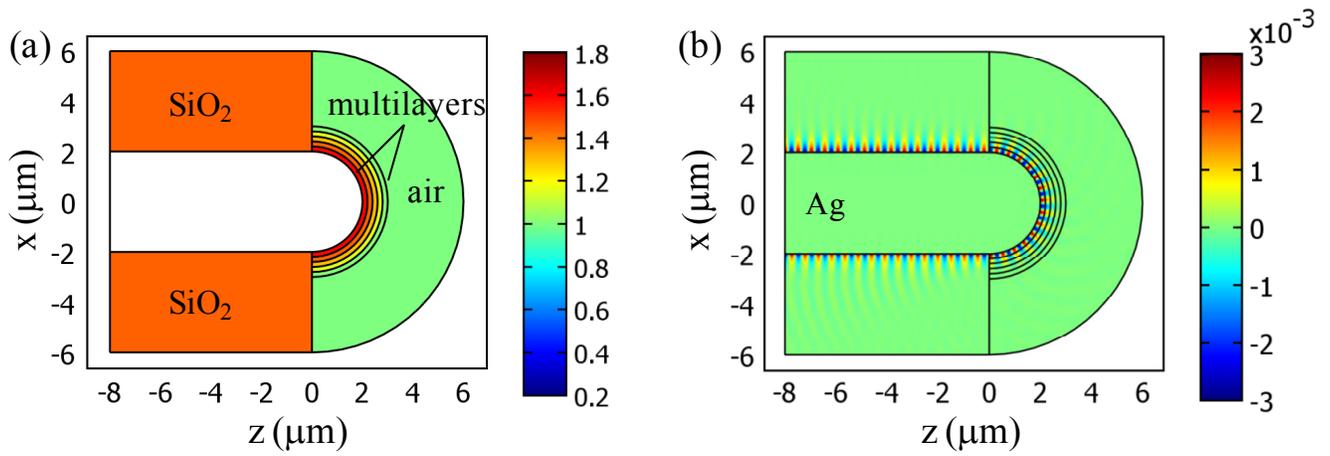

Figure 4. (a) The refractive indices in a realistic SPP bend design using a multilayer structure. The background refractive index in the straight region is 1.45 (silicon dioxide), while the transformed dielectric cladding in the 180° bend region is made of a multilayer with linearly changing refractive indices from 1.595 to 1.015. (c) The magnetic component of electromagnetic waves when SPPs propagate in the simplified bend structure, exhibiting a good transmission.

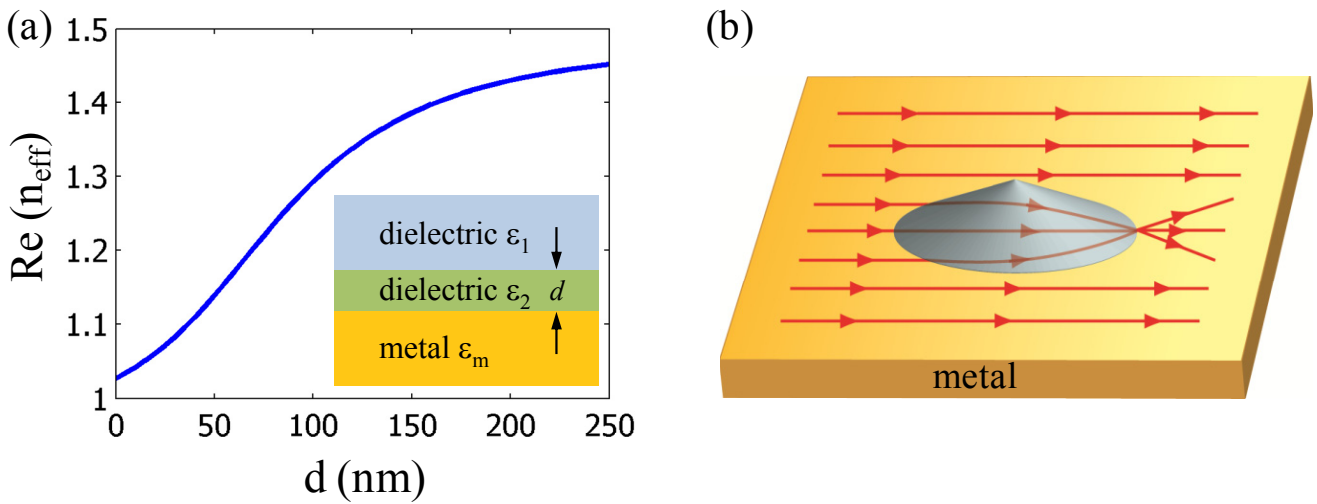

Figure 5. (a) Effective mode index of SPPs at 633 nm wavelength versus the thickness of the middle dielectric layer in a dielectric/dielectric/metal structure. The inset illustrates the structure. In the simulation, we choose $\varepsilon_1=1.0$, $\varepsilon_2=2.56$, and $\varepsilon_m=\varepsilon_{Ag}$. (b) The schematic of a plasmonic Luneburg lens, in which a dielectric cone is placed on a metal to focus SPPs.

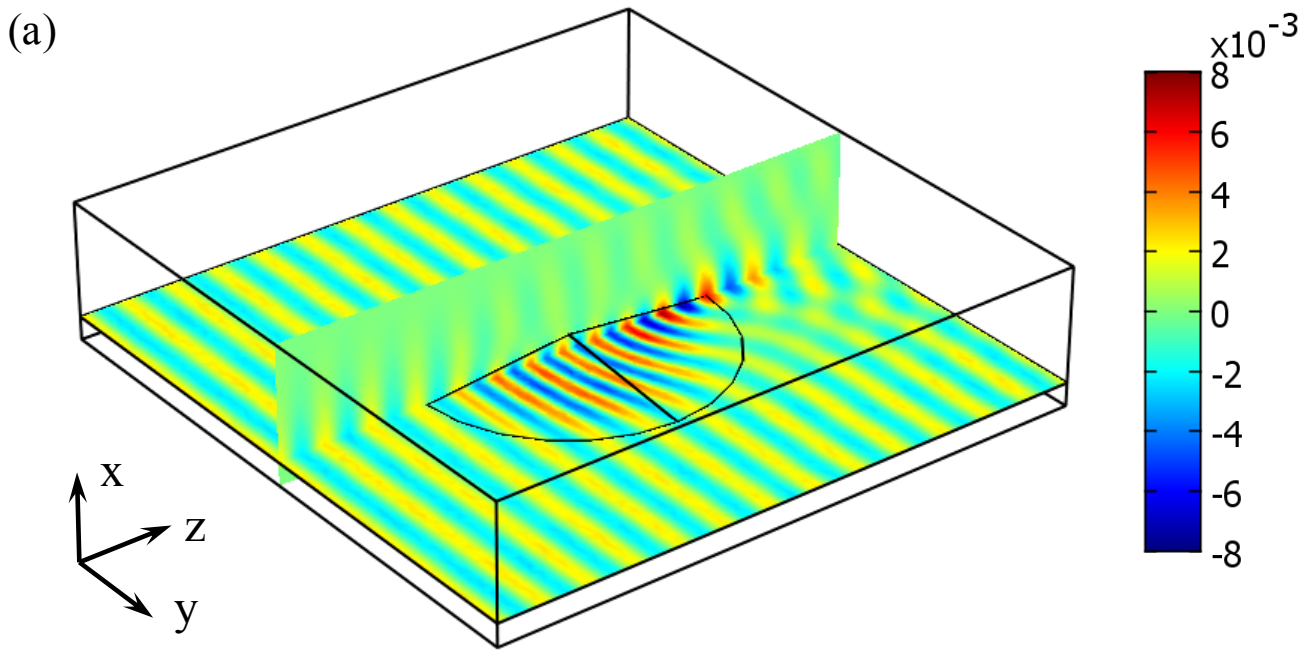

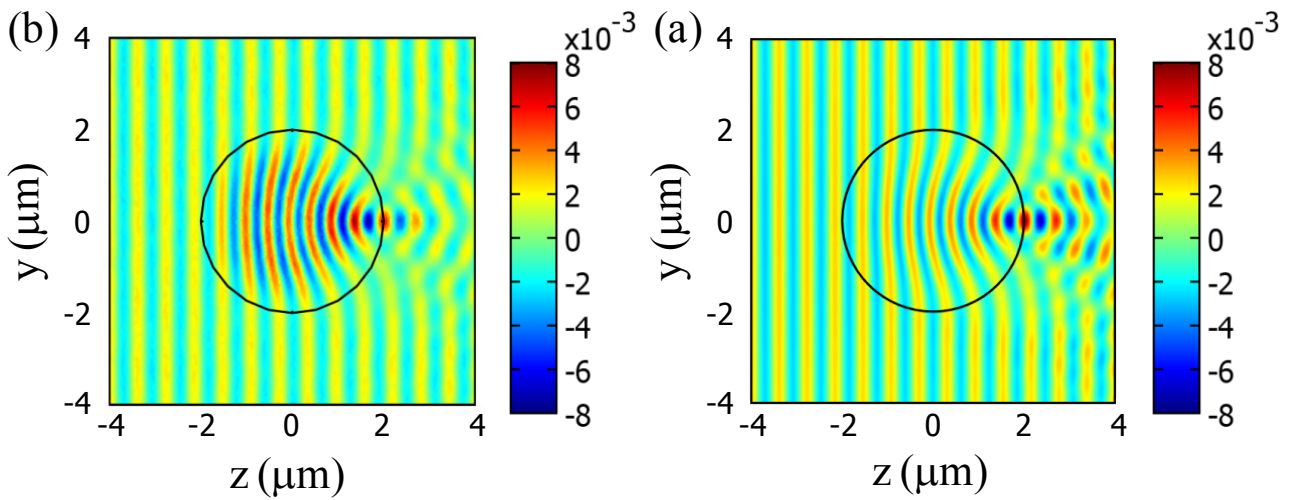

Figure 6. (a) A three-dimensional full-wave simulation for a plasmonic Luneburg lens, showing the magnetic component of the SPPs on two cross-sectional planes at 633 nm wavelength. SPPs are focused on the perimeter of the dielectric cone base. (b) The top view of the plasmonic Luneburg lens. The plotting plane is 5 nm above the silver surface. (c) A two-dimensional simulation of the Luneburg lens taking into account the effective mode index of SPPs.

REFERENCES

1. Ward, A. J.; Pendry, J. B. J. Mod. Opt. 1996, 43, 773-793.

2. Pendry, J. B.; Schurig, D.; Smith, D. R. Science 2006, 312, 1780-1782.

3. Leonhardt, U.; Philbin, T. G. New. J. Phys. 2006, 8, 247.

4. Leonhardt, U. Science 2006, 312, 1777-1780.

5. Smith, D. R.; Pendry, J. B.; Wiltshire, M. C. K. Science 2004, 305, 788-792.

6. Ramakrishna, S. A. Rep. Prog. Phys. 2005, 68, 449-521.

7. Schurig, D.; et al. Science 2006, 314, 977-980.

8. Chen, H. Y.; et al. Phys. Rev. Lett. 2009, 102, 183903.

9. Ma, Y. G.; Ong, C. K.; Tyc, T.; Leonhardt, U. Nature Mater. 2009, 8, 643-642.

10. Cai, W. S.; Chettiar, U. K.; Kildishev, A. V.; Shalaev, V. M. Nature Photon. 2007, 1, 224-227.

11. Rahm, M.; et al. Phys. Rev. Lett. 2008, 100, 063903.

12. Alù, A.; Engheta, N. Phys. Rev. E 2005, 72, 016623.

13. Smolyaninov, I. I.; Smolyaninova, V. N.; Kildishev, A. V.; Shalaev, V. M. Phys. Rev. Lett. 2009, 102, 213901.

14. Mei, Z. L.; Cui, T. J., Opt. Express 2009, 17, 18354-18363.

15. Li, J.; Pendry, J. B. Phys. Rev. Lett. 2008, 101, 203901.

16. Liu, R.; et al. Science 2009, 323, 366-369.

17. Valentine, J.; Li, J.; Zentgraf, T.; Bartal, G.; Zhang, X. Nature Mater. 2009, 8, 568-571.

18. Gabrielli, L. H., Cardenas, J., Poitras, C. B.; Lipson, M. Nature Photon. 2009, 43, 461-463.

19. Zentgraf, T., Valentine, J., Tapia, N., Li, J.; Zhang, X. Adv. Mater. (in press).

20. Kundtz, N.; Smith, D. R. Nature Mater. 2010, 9, 129-132.

21. H. Raether, Surface Plasmons: On Smooth and Rough Surfaces and on Gratings, Springer, Berlin, 1988.

22. Barnes, W. L.; Dereux, A.; Ebbesen, T. W. Nature 2003, 424, 824-830.

23. Drezet, A.; et al. Nano Lett. 2007, 7, 1697-1700.

24. Smolyaninov, I. I. New. J. Phys. 2008, 10, 115033.

25. Elser, J.; Podolskiy, V. A. Phys. Rev. Lett. 2008, 100, 066402.

26. Baumeier, B.; Leskova, T. A.; Maradudin, A. A.; Phys. Rev. Lett. 2009, 103, 246803.

27. Oulton, R. F.; Pile, D. F. P.; Liu, Y. M.; Zhang, X. Phys. Rev. B 2007, 76, 035408.

28. Johnson, P. B.; Christy, R. W. Phys. Rev. B 1972, 6, 4370-4379.

29. Hu, J.; Zhou, X. M.; Hu, G. K. Opt. Express 2009, 17, 1308-1320.

30. Chang, Z.; Hu, J.; Zhou, X. M.; Hu, G. K. http://arxiv.org/pdf/0912.4949.

31. Hasegawa, K.; Nöckel, J. U.; Deutsch M. Appl. Phys. Lett. 2004, 84, 1835-1837.

32. Pile, D. F. P.; Gramotnev, D. K. Opt. Lett. 2005, 30, 1186-1188.

33. Mei, Z. L.; Cui, T. J. J. Appl. Phys. 2009, 105, 104913.

34. Roberts, D. A.; Rahm, M.; Pendry, J. B.; Smith, D. R. Appl. Phys. Lett. 2008, 93, 251111.

35. Kwon, D.; Werner, D. H. New J. Phys. 2008, 10, 115023.

36. Luneburg, R. Mathematical Theory of Optics, Brown University, 1944.

37. Ditlbacher, H.; Krenn, J. R.; Schider, G.; Leitner, A.; Aussenegg, F. R. Appl. Phys. Lett. 2002, 81, 1762.

38. Huidobro, P. A.; Nesterov, M. L.; Martín-Moreno, L.; García-Vidal, F. J. submitted, 2010.